\documentclass[aps,prd,onecolumn,groupedaddress,showpacs,nofootinbib,amssymb]{revtex4}
\usepackage[dvips]{graphicx,color}
\usepackage{amssymb}
\usepackage{amsmath}
\usepackage{graphicx}
\usepackage{amsfonts}

\newcommand{\be}{\begin{equation}}
\newcommand{\ee}{\end{equation}}
\newcommand{\bea}{\begin{eqnarray}}
\newcommand{\eea}{\end{eqnarray}}
\newcommand{\beaa}{\begin{eqnarray*}}
\newcommand{\eeaa}{\end{eqnarray*}}

\begin{document}

\title{Loop Quantum Cosmology Gravitational Baryogenesis}
\author{
S.~D.~Odintsov,$^{1,2}$\,\thanks{odintsov@ieec.uab.es}
V.~K.~Oikonomou,$^{3,4}$\,\thanks{v.k.oikonomou1979@gmail.com}}
\affiliation{ $^{1)}$ICREA, Passeig Luis Companys,
23, 08010 Barcelona, Spain\\
$^{2)}$ Institute of Space Sciences (IEEC-CSIC)\\
C. Can Magrans s/n, 08193 Barcelona, SPAIN\\
$^{3)}$ Tomsk State Pedagogical University, 634061 Tomsk, Russia\\
$^{4)}$ Laboratory for Theoretical Cosmology, Tomsk State University of Control Systems
and Radioelectronics (TUSUR), 634050 Tomsk, Russia\\
}

\begin{abstract}
Loop Quantum Cosmology is an appealing quantum completion of classical cosmology, which brings along various theoretical features which in many cases offer remedy or modify various classical cosmology aspects. In this paper we address the gravitational baryogenesis mechanism in the context of Loop Quantum Cosmology. As we demonstrate, when Loop Quantum Cosmology effects are taken into account in the resulting Friedmann equations for a flat Friedmann-Robertson-Walker Universe, then even for a radiation dominated Universe, the predicted baryon-to-entropy ratio from the gravitational baryogenesis mechanism is non-zero, in contrast to the Einstein-Hilbert case, in which case the baryon-to-entropy ratio is zero. We also discuss various other cases apart from the radiation domination case, and we discuss how the baryon-to-entropy ratio is affected from the parameters of the quantum theory. In addition, we use illustrative exact solutions of Loop Quantum Cosmology and we investigate under which circumstances the baryon-to-entropy ratio can be compatible with the observational constraints.
\end{abstract}


\maketitle

\section{Introduction}

One of the toughest goals in theoretical physics is to find the quantum gravity theory that is responsible for all interactions in nature, and also will provide the most theoretically correct cosmological description for our Universe's evolution. This ultimate quantum gravitational cosmology will amend all the theoretical inconsistencies of modern cosmology, and also will provide a unified description for all the evolution eras of our Universe. In this research line, Loop Quantum Cosmology (LQC) \cite{LQC1,LQC3,LQC4,LQC5,LQC6,LQC19,LQC20} provides a quite appealing theoretical framework for cosmology, since many theoretical problems, like the occurrence of finite time singularities, find a consistent remedy in the context of LQC. For the most recent developments of LQC, see Refs. \cite{PAPLQC2,PAPLQC4,PAPLQC5,PAPLQC6,PAPLQC7,PAPLQC8,PAPLQC9,PAPLQC10,PAPLQC12,PAPLQC13,PAPLQC14,PAPLQC15,PAPLQC16,PAPLQC17,PAPLQC18,PAPLQC19,PAPLQC20,PAPLQC21,PAPLQC22,PAPLQC24,PAPLQC25,PAPLQC26}
 and references therein.

 In this paper the focus is on the gravitational baryogenesis mechanism in the context of LQC, and we aim to highlight the differences between the quantum and ordinary cosmology results. The gravitational baryogenesis mechanism \cite{Davoudiasl:2004gf,Lambiase:2006dq,Lambiase:2013haa,Lambiase:2006ft,Li:2004hh,
Pizza:2015epa,Odintsov:2016hgc,oikonsarid,oikonomoupan,newref,oikonomou} was proposed about a decade ago in Ref. \cite{Davoudiasl:2004gf}, in order to address one of the most fundamental unsolved problems in theoretical physics, which is the observed excess of matter over antimatter in the observable Universe. This excess is confirmed by Cosmic Microwave Background observations \cite{Bennett:2003bz} and also the Big Bang Nucleosynthesis predicts such an excess of matter over antimatter \cite{Burles:2000ju}. The gravitational baryogenesis mechanism is based on the presence of a $\mathcal{C}\mathcal{P}$-violating interaction term, and this is actually in accordance to one of the Sakharov criteria \cite{sakharov} which describe the necessary conditions for the generation of the baryon-anti-baryon asymmetry in the observed Universe. The baryon asymmetry gravitational baryogenesis term used in \cite{Davoudiasl:2004gf}, is of the form,
\begin{equation}
\label{baryonassterm}
\frac{1}{M_*^2}\int \mathrm{d}^4x\sqrt{-g}(\partial_{\mu} R) J^{\mu}\, ,
\end{equation}
and such a term can be justified by the presence of higher-order interactions which govern the fundamental
gravitational theory \cite{Davoudiasl:2004gf}. The parameter $M_*$ is the cutoff scale of the underlying effective gravitational theory, the current $J^{\mu}$ is the baryonic matter current and finally $g$ and $R$ are the determinant of the metric $g_{\mu \nu}$ and the Ricci scalar of the theory. In the case of Einstein-Hilbert gravity, if the background metric is a flat Friedmann-Robertson-Walker (FRW) metric, then the resulting baryon-to-entropy ratio $\eta_B/s$ in the case of a radiation dominated Universe, is zero, and therefore at the early stages of the Universe's evolution there is no baryon asymmetry generated by the gravitational baryogenesis mechanism. However, as we show in this work, if LQC effects are taken into account, the baryon-to-entropy ratio is non-zero, even in the radiation dominated case. We aim to explicitly find the general expression for the predicted baryon-to-entropy ratio in LQC, and also we investigate in detail the case for which the resulting baryon-to-entropy ratio can be compatible with the observational constraint $\eta_B/s<9 \times 10^{-11}$. To this end we shall consider various well known cosmologies resulting from LQC considerations, and we thoroughly analyze these, in the context of the gravitational baryogenesis mechanism.

This paper is organized as follows: In section II, we present the essential information on LQC, and we investigate how the gravitational baryogenesis mechanism is different in the context of LQC, in comparison to the Einstein-Hilbert case. In section III we study in detail the cases in which the resulting baryon-to-entropy ratio can be compatible with the observational data. Finally, the conclusions follow in the end of the paper.

\section{LQC Essentials and Gravitational Baryogenesis in LQC}

Before we proceed to the main goal of this paper, the calculation of the baryon-to-entropy ratio in the context of LQC, it is useful to recall some essential features of LQC. In the following we shall assume that the metric is a flat Friedmann-Robertson-Walker FRW metric, with line element,
\be
\label{metricfrw} ds^2 = - dt^2 + a(t)^2 \sum_{i=1,2,3}
\left(dx^i\right)^2\, ,
\ee
with $a(t)$ being the scale factor. Following the literature \cite{LQC1,LQC3,LQC4,LQC5,LQC6,LQC19,LQC20}, the theoretical framework of LQC is based on an effective Hamiltonian which describes the quantum aspects of the Universe, with the Hamiltonian being equal to,
\begin{equation}\label{effhamilt}
\mathcal{H}_{LQC}=-3V\frac{\sin^2(\lambda \beta)}{\gamma^2\lambda^2}+V\rho\, ,
\end{equation}
with $\gamma$, $\lambda$, and $V$ being the Barbero-Immirzi parameter, a parameter with dimensions of length and the volume $V=a(t)^3$, respectively, where $a(t)$ is the scale factor of the Universe. In addition, $\rho$ in Eq. (\ref{effhamilt}) denotes the energy density of the matter fluids present. In LQC, the Hamiltonian constraint $\mathcal{H}_{LQC}=0$, yields the following relation,
\begin{equation}\label{hamiltonianconstr}
\frac{\sin^2(\lambda \beta)}{\gamma^2\lambda^2}=\frac{\rho}{3}\, ,
\end{equation}
and moreover, by using the anti-commutation identity,
\begin{equation}\label{anticom}
\dot{V}=\{V,\mathcal{H}_{LQC}\}=-\frac{\gamma}{2}\frac{\partial \mathcal{H}_{LQC}}{\partial \beta},
\end{equation}
we arrive at the holonomy corrected Friedmann equation \cite{LQC1,LQC3,LQC4,LQC5,LQC6,LQC19,LQC20},
\begin{equation}\label{holcor1}
H^2=\frac{\kappa^2\rho}{3}\left (1-\frac{\rho}{\rho_c}\right )\, .
\end{equation}
The effective energy density satisfies the continuity equation,
\begin{equation}\label{cont}
\dot{\rho}(t)=-3H\Big{(}\rho(t)+P(t) \Big{)}\, ,
\end{equation}
with $P(t)$ being the total effective pressure of the matter fluid corresponding to the energy density $\rho$. By differentiating the holonomy corrected Friedman equation (\ref{holcor1}) with respect to the cosmic time $t$, and by using the continuity equation (\ref{cont}), we easily arrive at the following differential equation,
\begin{equation}\label{eqnm}
\dot{H}=-\frac{\kappa^2}{2}(\rho+P)(1-2\frac{\rho}{\rho_c})\, .
\end{equation}
The equations (\ref{holcor1}) and (\ref{eqnm}) will prove useful, since these can be combined to yield the Ricci scalar $R=12H^2+6\dot{H}$ in the flat FRW background. Before we proceed, we need to note that the LQC equations become identical to the classical equations if the limit $\rho_c\to \infty$ is taken, since the parameter $\rho_c$ quantifies the quantum effects of the theory. Indeed, in the limit $\rho_c\to \infty$, the equations (\ref{holcor1}) and (\ref{eqnm}) become equal to,
\begin{equation}\label{classicaleqns}
\dot{H}=-\frac{\kappa^2}{2}(\rho+P),\,\,\,H^2=\frac{\kappa^2\rho}{3}\, ,
\end{equation}
which are the classical equations of motion for an Einstein-Hilbert gravity in a FRW background.

The baryon-to-entropy ratio for the gravitational baryogenesis term of Eq. (\ref{baryonassterm}) reads \cite{Davoudiasl:2004gf},
\begin{equation}
\label{baryontoentropyrationori}
\frac{n_B}{s}\simeq -\frac{15g_b}{4\pi^2g_*}\frac{\dot{R}}{M_*^2
T}\Big{|}_{T_D}\, ,
\end{equation}
where $T_D$ is the decoupling temperature. Therefore a crucial element in the calculation of the baryon-to-entropy ratio in the context of gravitational baryogenesis, is to find the derivative of the Ricci scalar. In the following we shall assume that the matter content of the Universe is a perfect fluid with constant equation of state parameter $w$, pressure $P$ and energy density $\rho$, which are related as $P=w\rho$. In the Einstein-Hilbert case, the Ricci scalar can be easily deduced by using the Einstein equations, and it is equal to,
\begin{equation}\label{riccieinstein}
R=\kappa^2 \rho (1-3w)\, ,
\end{equation}
where $\kappa^2=\frac{1}{M_{p}^2}$ and $M_p$ is the Planck mass. The baryon-to-entropy ratio is determined by the derivative of the Ricci scalar $\dot{R}$, which in the Einstein-Hilbert case is,
\begin{equation}
\label{riccieinsteinhilbertder}
\dot{R}=\kappa^2 \dot{\rho} (1-3w),
\end{equation}
and hence it is easy to see that in the case of a radiation dominated Universe, in which case the equation of state parameter is $w=1/3$, the derivative of the Ricci scalar is zero, and therefore, the resulting baryon-to-entropy ratio is zero.

Let us now investigate the LQC case, and by combining Eqs. (\ref{holcor1}) and (\ref{eqnm}), the Ricci scalar $R=12 H^2+6\dot{H}$, reads,
\begin{equation}\label{riccisc1lqc}
R=\kappa^2\rho (1-3w)+2\frac{2\kappa^2\rho^2}{\rho_c}\left (1+3w\right )\, .
\end{equation}
Before we proceed let us discuss two important features of the LQC Ricci scalar (\ref{riccisc1lqc}), firstly that it contains the classical limit, which is quantified in the first term, since this is identical to that of Eq. (\ref{riccieinstein}). Secondly, in the limit $\rho_c\to \infty$, the second term in Eq. (\ref{riccisc1lqc}) vanishes, so the theory becomes classical. Now let us proceed to demonstrate the differences between the LQC and the classical case by comparing the resulting baryon-to-entropy ratio in the LQC case, which is determined by $\dot{R}$, which is,
\begin{equation}\label{baryonlqc}
\dot{R}=\kappa^2\dot{\rho} (1-3w)+4\frac{\kappa^2\rho \dot{\rho}}{\rho_c}\, .
\end{equation}
It can be easily seen from Eq. (\ref{baryonlqc}) that even for a radiation dominated Universe with $w=1/3$, the resulting baryon-to-entropy ratio is non-zero, and this is due to the existence of the second term in the LQC Ricci scalar of Eq. (\ref{riccisc1lqc}), which captures the quantum effects of the theory.

Hence, what we demonstrated so far is that the quantum holonomy effects of LQC, make the baryon-to-entropy ratio non-zero, a result which is different in comparison to the Einstein-Hilbert one, always in the context of the gravitational baryogenesis mechanism. It is conceivable that in the limit where the quantum effects vanish, that is when $\rho_c\to \infty$, the two cases yield a zero baryon-to-entropy ratio, as it is expected. In the next section we quantify our results by using several well known cosmologies of LQC, and we investigate under which circumstances the results can be compatible with the observational constraints.

\section{Examples of Gravitational Baryogenesis for LQC Cosmologies}

\begin{figure}[h]
\centering
\includegraphics[width=16pc]{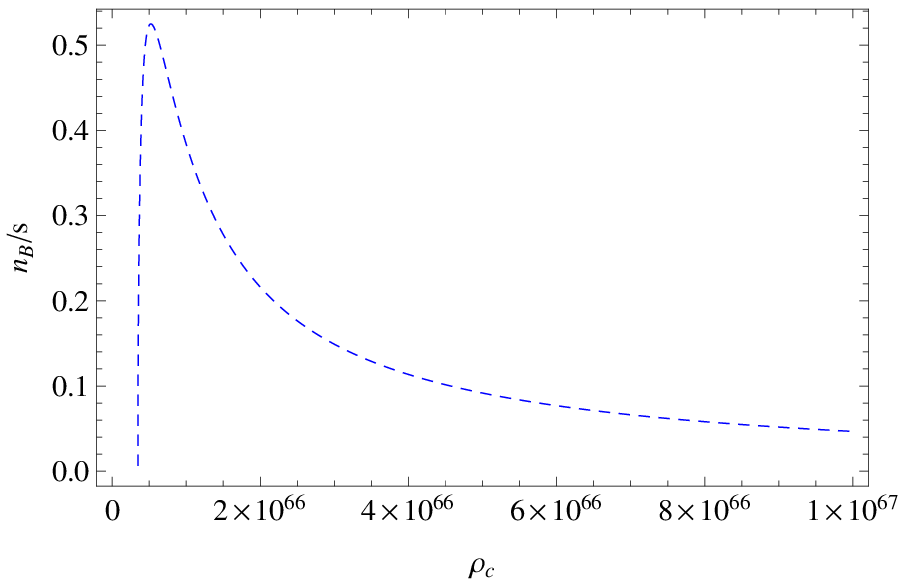}
\includegraphics[width=18pc]{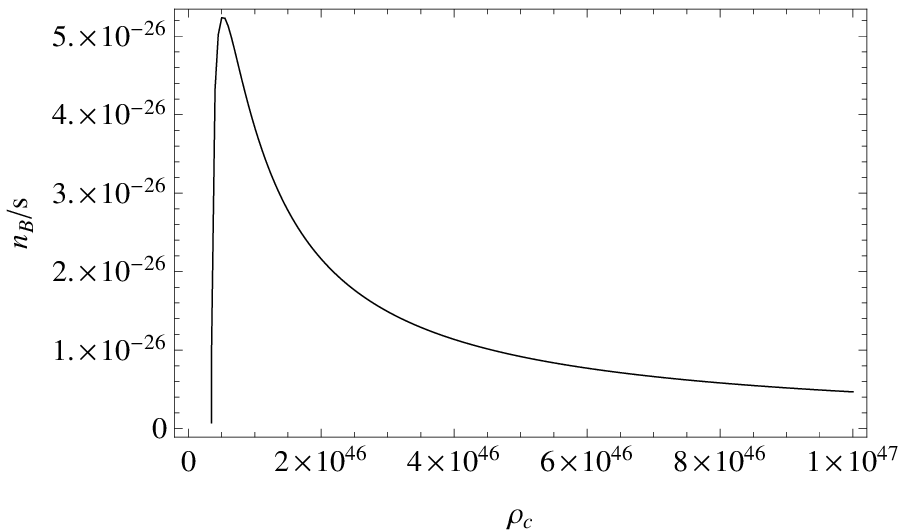}
\caption{{\it{The baryon-to-entropy ratio $n_B/s$ as a function of the parameter $\rho_c$ for $M_*=10^{12}$GeV, $g_b\simeq \mathcal{O}(1)$, $g_*=106$, for a radiation dominated Universe, so with $w=1/3$. The left plot corresponds to $T_D=2\times 10^{16}$GeV, while the right plot corresponds to $T_D=2\times 10^{11}$GeV.}}}
\label{plot2}
\end{figure}

In the context of LQC, the most well known examples of cosmologies are those corresponding to a perfect fluid with equation of state $P=w\rho$. In this case, by solving the equation of motion of LQC (\ref{holcor1}) in view of the continuity equation (\ref{cont}), with $P=w\rho$, the resulting scale factor is,
\begin{equation}\label{wscalefact}
a(t)=\left(\frac{3}{4} \kappa ^2(1+w)^2\rho_c t^2+1\right)^{\frac{1}{3 (1+w)}},
\end{equation}
while the corresponding Hubble rate is,
\begin{equation}\label{hubratemattfluid}
H(t)=\frac{t (1+w) \kappa ^2 \rho_c}{2 \left(1+\frac{3}{4} t^2 (1+w)^2 \kappa ^2 \rho_c\right)}\, .
\end{equation}
The cosmology described by the scale factor (\ref{wscalefact}) is very well known in the literature since it describes a non-singular bounce, see for example \cite{PAPLQC9,mattbounce1,mattbounce2,mattbounce3,mattbounce4,mattbounce5,mattbounce6} for the $w=0$ case, which is the matter bounce scenario, and also Refs. \cite{rad1,rad2} for the radiation bounce scenario. In addition, the energy density as a function of the cosmic time reads,
\begin{equation}\label{ernergydensimatt}
\rho (t)=\frac{4\rho_c}{4+3t^2(1+w)^2\kappa^2\rho_c}\, ,
\end{equation}
where we normalized the energy density to be equal to $\rho(0)=\rho_c$ at $t=0$. Assuming that the Universe evolves in a way so that it passes through states of thermal equilibrium (quasi-static thermal equilibrium), the total energy density as a function of the temperature is,
\begin{equation}
\label{equilibrium}
\rho=\frac{\pi^2}{30}g_*\, {T}^4\, .
\end{equation}
We can easily find the decoupling time $t_D$ as a function of the decoupling temperature $T_D$, by combining Eqs. (\ref{ernergydensimatt}) and (\ref{equilibrium}), and the result is,
\begin{equation}\label{dectime}
t_D=\frac{2}{\sqrt{3} \pi } \sqrt{\frac{30 \rho_c-\pi ^2 g_* T_D^4}{g_* \kappa ^2 \rho_c T_D^4 (w+1)^2}}\, .
\end{equation}
From Eq. (\ref{dectime}) we can see that the parameter $\rho_c$ is constrained to satisfy the inequality $30 \rho_c>\pi ^2 g_* T_D^4$, so it must be at least four orders larger in comparison to the decoupling temperature. By using Eq. (\ref{ernergydensimatt}) and substituting in Eq. (\ref{baryonlqc}), we find that the term $\dot{R}$ as a function of the cosmic time is,
\begin{equation}\label{dotr2}
\dot{R}=\frac{4 \kappa ^2 \rho_c \left((1-3 w) \left(3 \kappa ^2 \rho_c t^2 (w+1)^2+4\right)^2-96 \kappa ^2 \rho_c t (w+1)^2 (3 w+1)\right)}{\left(3 \kappa ^2 \rho_c t^2 (w+1)^2+4\right)^3}\, ,
\end{equation}
so by using Eq. (\ref{dectime}), the term $\dot{R}$ as a function of the decoupling temperature is,
\begin{equation}\label{dotr3}
\dot{R}=-\frac{\pi ^2 g_* \kappa ^2 T_D^4 \left(\pi ^3 g_*^2 \kappa ^2 T_D^8 (w+1)^2 (3 w+1) \sqrt{\frac{90 \rho_c-3 \pi ^2 g_* T_D^4}{g_* \kappa ^2 \rho_c T_D^4 (w+1)^2}}+225 \rho_c (3 w-1)\right)}{6750 \rho_c}\, .
\end{equation}
Notice the minus sign appearing in Eq. (\ref{dotr3}) which is important since the resulting baryon-to-entropy (\ref{baryontoentropyrationori}) should be positive. By substituting $\dot{R}$ from Eq. (\ref{dotr3}) in Eq. (\ref{baryontoentropyrationori}), the final expression for the baryon-to-entropy ratio of a LQC perfect fluid with $P=w\rho$, is equal to,
\begin{equation}\label{resultbaryontoentropy}
\frac{n_B}{s}\simeq \frac{g_b \kappa ^2 T_D^3 \left(\pi ^3 g_*^2 \kappa ^2 T_D^8 (w+1)^2 (3 w+1) \sqrt{\frac{90 \rho_c-3 \pi ^2 g_* T_D^4}{g_* \kappa ^2 \rho_c T_D^4 (w+1)^2}}+225 \rho_c (3 w-1)\right)}{1800 M_*^2 \rho_c}\, .
\end{equation}
Let us now investigate under which circumstances the resulting baryon to entropy ratio can be compatible with the theoretical bound $n_B/s\preceq 9 \times 10^{-11}$. We shall assume Planck units for simplicity, and we choose the cutoff scale $M_*$ is equal to $M_*=10^{12}$GeV, and also that the decoupling temperature is $T_D=M_I=2\times 10^{16}$GeV, where $M_I$ is the upper
bound for tensor-mode fluctuations constraints on the inflationary scale.
\begin{figure}[h]
\centering
\includegraphics[width=16pc]{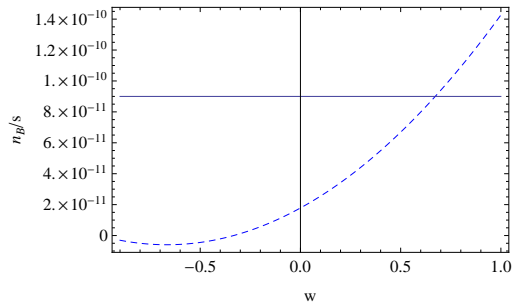}
\caption{{\it{The baryon-to-entropy ratio $n_B/s$ as a function of the equation of state parameter $w$ for $M_*=10^{12}$GeV, $T_D=2\times 10^{16}$GeV, $g_b\simeq \mathcal{O}(1)$, $\rho_c\simeq 10^{76}$GeV$^4$, $g_*=106$, with $w$ taking values $-0.9\leq w\leq 1$.}}}
\label{plot1}
\end{figure}
Also we assume that, $g_b\simeq \mathcal{O}(1)$ and also that $g_*=106$ which is the total number of the effectively massless particles in the early Universe. Finally, we assume that $\rho_c\simeq 10^{76}$GeV$^4$ and we can vary the effective equation of state parameter $w$, to obtain various interesting limiting cases. For example, when $w=1/3$, we have the radiation dominated Universe, in which case the baryon-to-entropy ratio predicted by gravitational baryogenesis (\ref{resultbaryontoentropy}) becomes approximately equal to $n_B/s\simeq 4.75583\times 10^{-11}$, which is compatible with the observational bounds. Also by choosing $w=0$, which corresponds to the matter dominated epoch, the resulting baryon-to-entropy ratio is $n_B/s\simeq 1.78331\times 10^{-11}$, which is again compatible with the observational bounds. Consider now a quintessential fluid with $w<0$. It can be shown that when $w$ is approximately smaller than $w\preceq -0.335$, the resulting baryon-to-entropy ratio is negative, therefore these cases have no physical interest. In Fig. \ref{plot1}, we plot the baryon-to-entropy ratio $n_B/s$ as a function of the effective equation of state parameter $w$, for $M_*=10^{12}$GeV, $T_D=2\times 10^{16}$GeV, $g_b\simeq \mathcal{O}(1)$, $\rho_c\simeq 10^{76}$GeV$^4$, $g_*=106$, with $w$ taking values $-0.9\leq w\leq 1$. The horizontal line corresponds to the observational bound $9\times 10^{-11}$. It can be seen that depending on the value of $w$, the results are quite different, other being un-physical, and other results compatible with observational constraints. Before closing we need to note that the resulting baryon-to-entropy ratio crucially depends on the parameter $\rho_c$, which for consistency has to be roughly larger than the fourth power of the decoupling temperature, that is $\rho_c>T_D^4$. In order to see how the baryon-to-entropy ratio behaves as a function of the critical density parameter $\rho_c$, in Fig. \ref{plot2}
we plotted the behavior of the baryon-to-entropy ratio as a function of $\rho_c$, for $M_*=10^{12}$GeV, $g_b\simeq \mathcal{O}(1)$, $g_*=106$, for a radiation dominated Universe, so for $w=1/3$. The left plot corresponds to $T_D=2\times 10^{16}$GeV, while the right plot corresponds to $T_D=2\times 10^{11}$GeV. As it can be seen, as the decoupling temperature gets lower values, the baryon-to-entropy ratio gets the observationally accepted values, for lower values of the parameter $\rho_c$.

\section{Conclusions}

In this paper we investigated the effects of LQC originating holonomy corrections on the gravitational baryogenesis mechanism. Particularly, we calculated the baryon-to-entropy ratio for a LQC governed FRW Universe, in the context of the gravitational baryogenesis mechanism, in which case the term $\dot{R}$ is involved in the calculation. The most sound result of our work is the fact that even in the case of a radiation perfect fluid controlling the evolution, the baryon-to-entropy ratio is non-zero, which is in contrast to the Einstein-Hilbert case, where the baryon-to-entropy ratio is zero. Therefore, if the gravitational baryogenesis mechanism is considered a viable baryon asymmetry generating mechanism, the LQC effects drastically affect the amount of baryon asymmetry in the early Universe. We used quite well-known cosmological evolutions which are solutions to the holonomy corrected FRW equations in order to quantify our results and to investigate when concordance with the observational bounds can be achieved. Particularly, we used several bouncing cosmologies, like the matter bounce and the radiation bounce, and we examined under which conditions the observational bounds are respected. As we showed, the observational bounds on the baryon-to-entropy ratio are well respected by suitably adjusting the parameters, and as a comment we need to note that in order for the resulting picture to be considered viable, the critical density parameter must be larger than the fourth power of the decoupling temperature, that is $\rho_c\succeq T_D^4$.

Finally, let us note that we used the most frequent holonomy-corrected FRW equations existing in the literature, however there exist various non-trivial extensions of the holonomy effects we used for our study, like for example the ones appearing in Ref. \cite{LQC20}. In principle one could add a thorough study on these cases too, but we expect that the qualitative results will be the same, so a non-zero baryon-to-entropy ratio should occur in these cases too.

\section*{Acknowledgments}

This work is supported by MINECO (Spain), project
 FIS2013-44881 (S.D.O) and by Min. of Education and Science of Russia (S.D.O
and V.K.O).

\end{document}